\documentclass[%
 reprint,
 amsmath,amssymb,
 aps,
]{revtex4-2}

\usepackage[T1]{fontenc}
\usepackage{xcolor}%
\usepackage{graphicx}
\usepackage{dcolumn}
\usepackage{bm}
\usepackage{mathtools} 
\usepackage{stmaryrd}

\newcommand{\Ivec}{\mathbf{I}}
\newcommand{\Qvec}{\mathbf{Q}}

\graphicspath{{Pictures/}} 

\begin{document}

\preprint{APS/123-QED}

\title{Tensorial Model of Chiral Smectic C Liquid Crystals}
\author{Jingmin Xia}
\affiliation{College of Meteorology and Oceanography, National University of Defense Technology, Changsha 410072, China}
\author{Jinbing Wu}
\affiliation{National Laboratory of Solid State Microstructures, Jiangsu Physical Science Research Center, College of Engineering and Applied Sciences, Nanjing University, Nanjing 210023, China}
\author{Yucen Han}
\email{yucen.han@ruc.edu.cn}
\affiliation{Center for Applied Mathematics, Renmin University of China 100872, Beijing, China}
\begin{abstract}
We propose a continuum tensorial model for chiral smectic C (SmC$^*$) liquid crystals using a tensor-valued order parameter $\Qvec$ to describe orientational order and a real-valued order parameter $\delta\rho$ to capture layer modulation. This model accounts for the coupled effects of nematic alignment, smectic layering, chirality and spontaneous polarisation inherent in SmC$^*$ systems. 
The model is
validated through a series of numerical experiments --- helix, bistable switching, bookshelf and chevron, and defect in SmC$^*$. 
Our results highlight the model's capability to describe complex phenomena in SmC$^*$ liquid crystals, providing a robust foundation for further theoretical and applied studies on phase transitions. 
\end{abstract}
\maketitle

\section{Introduction}
\begin{figure}
    \includegraphics[width=1.0\columnwidth]{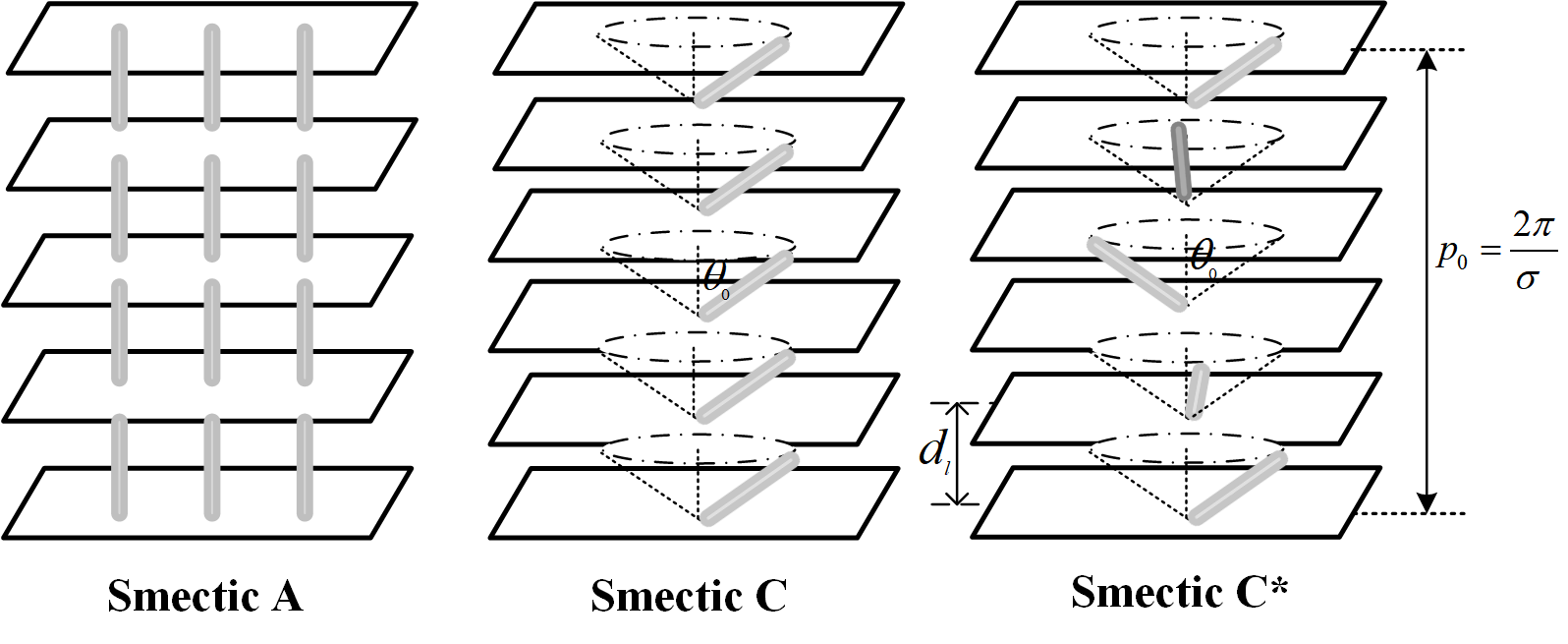}
    \caption{Schematic illustration of SmA, SmC and SmC$^*$ phases, where molecules form layered structures. In the SmA phase, molecules are perpendicular to layers, while SmC molecules tilt from layer normals with the angle $\theta_0\neq 0$. In the SmC$^*$ phase, molecules retain the tilt angle as in the SmC phase while rotating helically with the pitch $p_0$ on the cone surface, leading to the chirality. $d_{\ell}$ is the layer thickness of SmC$^*$.
   }
    \label{fig:phases}
\end{figure}

The chiral smectic C (SmC$^{*}$) phase is first reported in the mid-1970s by Meyer et al.~\cite{Meyer1975}. The SmC$^{*}$ system is characterised by a helical structure, where the molecules tilt with respect to the smectic layer normal and rotate on the cone surface of the tilt angle. The intrinsic helical superstructure of the phase distinguishes SmC$^{*}$ sharply from smectic C (SmC), where molecules are tilted but lack chirality, and smectic A (SmA), in which the molecules align with the layer normal, see Fig.~\ref{fig:phases}.
The absence of mirror symmetry leads to the emergence of a spontaneous polarisation within layers. It gives rise to the field of ferroelectric liquid crystals (FLCs), which is one of the most significant discoveries in liquid crystal science \cite{gennes-book, stewart-2004-book, yoshizawa2024}. 
Owing to its ferroelectric switching behaviour, enabling fast electro-optic response and bistability \cite{1989nature, gennes-book, song2024giant}, SmC$^{*}$ is highly valuable for applications in ferroelectric liquid crystal displays (FLCDs)~\cite{ClarkLagerwall1980}. Beyond display technologies, current research continues to explore SmC$^{*}$ for advanced photonic and nonlinear optical applications \cite{tsuji-2012-article, chen-2024NSR}, where its layered architecture, chirality, and polarisation coupling offer unique opportunities.

Mathematical modelling plays a crucial role in understanding and predicting the behaviour of ferroelectric SmC$^*$ liquid crystals (LCs). Although the existing mathematical models of SmC$^*$ phases are capable of capturing the helical structure and their response to external electric fields, they remain mostly limited to theoretical analysis \cite{2007joo} or one-dimensional numerical studies \cite{park2010modeling}. To simplify the theoretical work, researchers directly assume that the layer normal coincides with the $z$-axis \cite{lagerwall2004ferroelectric}. Some SmC$^*$ works introduce a real-valued function $u$ to represent the vertical displacement of layers \cite{stewart2019static}, which restricts the analysis to cases with only small layer deformations. An alternative formulation employs the complex-valued smectic order parameter \cite{chen1976landau}, but from a computational perspective, the use of complex order parameters can lead to numerical difficulties and even potential physical inconsistencies when interpreting the imaginary component \cite{pevnyi-2014-article}.
The director field is always described by a unit vector field $\mathbf{n}$, and its distortions are described using the Oseen--Frank energy (splay, twist, and bend) \cite{frank1958liquid}. Nevertheless, such a vector-based representation does not respect the head-to-tail symmetry of liquid crystal molecules, cannot capture half-integer defects, and lacks the ability to describe the biaxiality of LCs\cite{ball2017liquid,han2021solution}.
To enable numerical computation in higher-dimensional settings with complex geometrical confinement, and to investigate SmC$^*$ structures with defects, we propose in this manuscript a tensorial model to overcome the challenges mentioned above.

In previous works, the authors present tensorial models to describe SmA \cite{xia-2021-article} and SmC \cite{xia2024prr} phases.
The analysis and numerical results have demonstrated both theoretical robustness and qualitative agreement with other modelling methodologies and experimental observations \cite{shi2025modified, xia-2023b-article, xia-2021-article}.
Based on these works, we present a tensorial model to describe ferroelectric chiral SmC LCs, using a real-valued smectic order parameter $\delta\rho$ representing the variation from the average density, and a tensor-valued nematic order parameter $\Qvec$ from Landau--de Gennes theory \cite{gennes-book}. Three essential model parameters for SmC$^*$ are: $\sigma$ giving the pitch of the helix, $\theta_0$ representing the tilt angle of molecules from the smectic layer normal, and $q$ corresponding to the layer spacing. By employing the finite element method to compute the minimisers of the proposed energy functional, we investigate the classical structures and behaviors of the SmC$^{*}$ phase, such as the helix structure, bistable switching induced by external electric fields, the occurrence of bookshelf and chevron layer geometries in SmC$^{*}$, as well as the coupling between defects and the helix. These numerical results confirm the validity and predictive capability of our proposed model.

\section{Mathematical model}

In the absence of surface energies, the free energy for SmC$^*$ LCs is given by
\begin{equation}\label{eq:energy_LdG}
\begin{aligned}
    F(\Qvec,\delta\rho)&=\int_{\Omega} \bigg\{
    f_{el}(\Qvec)+f_{bQ}(\Qvec)+f_{b\delta\rho}(\delta\rho) \\
    &~+ f_{layer}(\delta\rho)+ f_{angle}(\Qvec,\delta\rho) 
    \bigg\}\mathrm{d}V,
\end{aligned}
\end{equation}
where $\Qvec$ is a tensor-valued order parameter which is a macroscopic measure of the liquid crystal anisotropy or directionality in the famous Landau--de Gennes theory (LdG, readers may refer to \cite[Sections 2.1, 2.3 and 3.1]{gennes-book}). 
Specifically, the order parameter $\Qvec$ is a symmetric traceless $3\times 3$ matrix. From the spectral decomposition theorem, $\mathbf{Q}$ can be written as $\mathbf{Q} = \sum_{i = 1}^3\lambda_i\mathbf{n}_i\otimes\mathbf{n}_i$, where $\mathbf{n}_i$, $i = 1,2,3$ are the eigenvectors of $\mathbf{Q}$ and $\lambda_i$, $i = 1,2,3$ are the corresponding eigenvalues, respectively. The eigenvectors represent the preferred orientations of the spatially averaged local molecular alignment, i.e., the nematic directors, while the corresponding eigenvalues quantify the degree of orientational order along these directions. 
The density variation $\delta\rho = \rho-\rho_0$ represents the density deviation from the average density value $\rho_0$. We regard the variation of density from low to high and back to low --- namely, $\delta\rho$ changing from minimum to maximum and then to minimum again --- as one smectic layer.
The domain can take $\Omega = [0,L]\times[0,W]\times[0,H]$, for instance.

The first term in \eqref{eq:energy_LdG} is the elastic energy density, which comes from the elastic energy density of cholesteric phase in \cite{fukuda2010cholesteric}, formulated as
\begin{equation}\label{eq:Q-elastic}
f_{el}(\Qvec) = \frac{K_0}{4}(\nabla\cdot\Qvec)^2+\frac{K_1}{4}|\nabla\times\Qvec+2\sigma \Qvec|^2,
\end{equation}
where $K_0>0$ and $K_1>0$ are material-dependent twist and splay/bend elastic constants, respectively. 
The parameter $\sigma=2\pi/p_0$ corresponds to the intrinsic helical wavenumber. The parameter $p_0$ is the pitch of the helical structure.  When $\sigma = 0$, this term reduces to the conventional nematic elastic form \cite{gennes-book}, and can describe achiral smectic liquid crystal phases, such as SmA and SmC.
Here,
$\nabla\cdot\Qvec=\partial_\alpha Q_{i\alpha}$ and $\nabla\times\Qvec=\epsilon_{i,j,k}\partial_j Q_{k\beta}$, where $\epsilon_{i,j,k}$ is the Levi--Civita anti-symmetric symbol.
 
The nematic bulk energy density is denoted as the classical LdG form
\begin{equation}\label{eq:bulk_Q_LdG}
    f_{bQ}(\Qvec):=\frac{A}{2}\textrm{tr}\Qvec^2-\frac{B}{3}\textrm{tr}\Qvec^3+\frac{C}{4}(\textrm{tr}\Qvec^2)^2,
\end{equation}
with coefficients $A$, $B$, and $C$ in the LdG expansion of the free energy.
The parameter $A=\alpha (T  - T^*)$ is the rescaled temperature, with $\alpha  > 0$, and $T^*$ is the characteristic liquid crystal temperature. For $A<0$, the minimisers of $f_{bQ}$ constitute a continuum of uniaxial $\Qvec$-tensor, i.e., $\left\{\Qvec = s^*(\mathbf{n}\otimes\mathbf{n}-\mathbf{I}/3)\right\}$ where $s^* = \frac{B+\sqrt{B^2+24|A|C}}{4C}$. $B>0$ and $C>0$ are material-dependent constants. The reported values
for N-(4-methoxybenzylidene)-4-butylaniline (MBBA) as a
representative nematic material are $B = 0.64 \times 10^4N/m^2$ and
$C = 0.35 \times 10^4 N/m^2$.

The smectic bulk energy density of the density variation $\delta\rho$ is similarly derived as
\begin{equation}\label{eq:bulk_deltarho_LdG}
    f_{b\delta\rho}(\delta\rho):=\frac{D}{2}(\delta\rho)^2-\frac{E}{3}(\delta\rho)^3+\frac{F}{4}(\delta\rho)^4,
\end{equation}
where $D$, $E$ and $F$ correspond to temperature-, concentration- and material-dependent constants, respectively. 

To ensure the equi-distant layer structures in smectics, the following energy density is adopted
\begin{equation}
f_{layer}(\delta\rho) = \Lambda_1\left(\Delta\delta\rho + q^2\delta\rho\right)^2,
\end{equation}
where $q$ represents the wave number of smectics. If we assume $\delta\rho = \cos(\mathbf{k}\cdot\mathbf{x})$, where $\mathbf{x}$ is the spatial coordinate, we have $(\Delta\delta\rho + q^2\delta\rho)^2 = (-|\mathbf{k}|^2\delta\rho + q^2\delta\rho)^2$, which attains its minimiser when $|\mathbf{k}| = q$. The constant $\Lambda_1$ is the weight for penalising deviations from equidistant layer spacing. Its magnitude scales on the order of $\mathcal{O}(q^{-4})$ and carries units of $N\cdot m^2$. The wave number $q$ depends on the layer thickness $d_{\ell}$, i.e., $q =  \frac{2\pi}{d_{\ell}}$. 

The last term in \eqref{eq:energy_LdG} imposes the tilt angle $\theta_0$ between the directors and layer normals
\begin{equation}
f_{angle}(\Qvec,\delta\rho) = 
\Lambda_2 \left(\mathrm{tr}(\mathcal{D}^2\delta\rho(\mathbf{Q}+\frac{\Ivec}{3})) + q^2\delta\rho\cos^2\theta_0\right)^2,
\end{equation}
where $\mathcal{D}^2$ denotes the Hessian operator, and $\theta_0$ is the preferred angle between the layer normal and the director. 
We assume a density modulation of the form 
$
\delta \rho = \cos(\mathbf{k} \cdot \mathbf{x}),
$
where $\mathbf{k} = q\mathbf{m}$, and $\mathbf{m}$ denotes the unit layer normal.  
In this case, one obtains
$
\mathcal{D}^2 \delta \rho = -q^2\delta \rho (\mathbf{m} \otimes \mathbf{m}).
$
Furthermore, if the order parameter $\Qvec$ is taken to be uniaxial with director $\mathbf{n}$, i.e.,
$
\Qvec = s^*(\mathbf{n} \otimes \mathbf{n} - \mathbf{I}/3),
$
the subsequent evaluation shows that the coupling term is minimised when
$
\mathbf{m} \cdot \mathbf{n} = \cos \theta_0,
$
meaning that $f_{angle}$ penalizes the tilt angle between the director $\mathbf{n}$ and layer normal $\mathbf{m}$ to be $\theta_0$. The constant $\Lambda_{2}$ serves as the weighting factor for this coupling effect, 
with its magnitude scaling as $\mathcal{O}(q^{-4})$ and carrying units of $N\cdot m^{2}$. 

The decyloxybenzylidene amino 2-methyl butyl cinnamate (DOBAMBC) served as the working compound in the pioneering study of Clark and Lagerwall (1980), which first demonstrated ferroelectric switching in LCs. The typical values of the three essential model parameters for DOBAMBC as a
representative SmC$^*$ material are $d_{\ell}=3.2nm$, $\theta_0=20^\circ
\sim 25^\circ$, and $p_0=0.3\mu m\sim 1\mu m$ \cite{ClarkLagerwall1980,lagerwall2004ferroelectric}. In Sec. \ref{sec:results}, the parameters chosen in the numerical simulations are based on phenomenological considerations.

By rescaling the system according to $\bar{x} = \frac{x}{H}$, $\bar{y} = \frac{y}{H}$, $\bar{z} = \frac{z}{H}$, the non-dimesionalised free energy is given by
\begin{equation}\label{eq:energy-3d}
\begin{aligned}
    &\bar{F}(\Qvec,\delta\rho) = \frac{F(\Qvec,\delta\rho)}{K_0 H}  \\
    &= \int_{\bar{\Omega}}\bigg\{
\frac{1}{4}(\bar{\nabla}\cdot\Qvec)^2+\frac{\eta}{4}|\bar{\nabla}\times\Qvec+2\sigma_0 \Qvec|^2\\
&~+ \frac{a}{2}\textrm{tr}\Qvec^2-\frac{b}{3}\textrm{tr}\Qvec^3+\frac{c}{4}(\textrm{tr}\Qvec^2)^2\\
&~ + \frac{d}{2}(\delta\rho)^2-\frac{e}{3}(\delta\rho)^3+\frac{f}{4}(\delta\rho)^4 + \lambda_1\left(\bar{\Delta}\delta\rho + q_0^2\delta\rho\right)^2\\
&~ + \lambda_2 \left(\mathrm{tr}(\bar{\mathcal{D}}^2\delta\rho(\mathbf{Q}+\frac{\Ivec}{3})) + q_0^2\delta\rho\cos^2\theta_0\right)^2 \bigg\}\textrm{d}\bar{V},
\end{aligned}
\end{equation}
where $\bar{\Omega} = [0,l]\times[0,w]\times[0,1]$, $
\Ivec$ denotes the $3\times 3$ identity matrix. 
The non-dimensionalised parameters are given by
\begin{gather}
l = \frac{L}{H},\ w = \frac{W}{H},\ \eta = \frac{K_1}{K_0},\ \sigma_0 = \sigma H,\ a = \frac{AH^2}{K_0},\nonumber\\
b = \frac{BH^2}{K_0},\ c = \frac{CH^2}{K_0},\ d = \frac{DH^2}{K_0},\ e = \frac{EH^2}{K_0},\nonumber\\
f = \frac{FH^2}{K_0},\ q_0 = qH,\ \lambda_1 = \frac{\Lambda_1}{K_0H^2},\ \lambda_2 = \frac{\Lambda_2}{K_0H^2}.\nonumber
\end{gather}

From now on, we drop bars for brevity.
\section{Numerical method}
The finite element method is employed in our numerical computations.
Specifically, we divide the unit interval $ [0,1]$ into 100 equi-distanced pieces in Sec.~\ref{sec:results}(A-B).
We discretise the square $[0,1]^2$ in Sec.~\ref{sec:results}(C) into $30\times 30$ pieces and further triangulated in both diagonal directions.
The unit disc in Sec.~\ref{sec:results}(D) is unstructured triangulated via Gmsh \cite{gmsh} and imported in Firedrake \cite{firedrake} as the mesh. We employ piecewise quadratic functions for approximating the components of the symmetric traceless tensor $\Qvec$ and piecewise cubic functions for representing the density variable $\delta\rho$.
Moreover, due to the same reason for the higher regularity of $\delta\rho$ as discussed in \cite{xia-2021-article}, we use the $\mathcal{C}^0$ interior penalty approach throughout our paper, where the mesh scale $h_e$ is selected as the average of the diameters of the cells located on either side of an edge or face.

Since the total energy under minimisation is nonlinear, we employ Newton's method with an $L^2$ line-search as the outer nonlinear solver, see \cite[Algorithm 2]{brune2015} for further details. The nonlinear solver is considered to have converged when the Euclidean norm of the residual falls below $10^{-8}$ or decreases by a factor of $10^{-16}$ from its initial value, whichever occurs first.
For the inner solves, the linearised systems are tackled using the sparse LU factorization library MUMPS \cite{mumps}. This solver framework is implemented within the Firedrake library \cite{firedrake}, which relies on PETSc \cite{petsc} for solving the resulting linear systems.

\section{Results and discussions}\label{sec:results}
We perform the following numerical experiments to validate our proposed tensorial model for SmC$^*$ phase as stated in \eqref{eq:energy_LdG}. 

\textit{(A) Helix}---Firstly, to study the basic helical structure in the SmC$^*$ phase, we restrict ourselves to a one-dimensional geometry: $\Omega = [0, 1]$, assuming the profile is invariant in the other two dimensions.
We impose the following Dirichlet boundary conditions on $x=0$ and $x=1$:
\begin{equation*}
    \Qvec(0)=\Qvec(1) = s^*(\mathbf{n}_b\otimes\mathbf{n}_b-\Ivec/3),
\end{equation*}
where $\mathbf{n}_b = [\cos\theta_0,0,\sin\theta_0]^\top$
and 
\begin{equation}\label{eq:s*}
s^*=\frac{b+\sqrt{b^2+24|a|c}}{4c},
\end{equation}
is the uniaxial order parameter that minimises the bulk energy density of $\mathbf{Q}$ in \eqref{eq:energy-3d}. 
Notice that we impose no specific boundary condition on the density variation $\delta\rho$ as we did in our previous SmA and SmC work in \cite{xia2024prr,xia-2021-article}. The
initial guess of $\mathbf{Q}$ is $\mathbf{Q}_0 = s^*(\mathbf{n}_0\otimes\mathbf{n}_0-\Ivec/3)$ where the director $\mathbf{n}_0$ is $$\mathbf{n}_0=\left[\cos(\theta_0), \sin(\theta_0)\sin(-\sigma_0 x), \sin(\theta_0)\cos(-\sigma_0 x)\right]^\top.$$
Meanwhile, the initial guess of $\delta\rho$ is simply a sinusoidal function $ \delta\rho_0 = \sin(q x)$.

\begin{figure}
    \includegraphics[width=1.0\columnwidth]{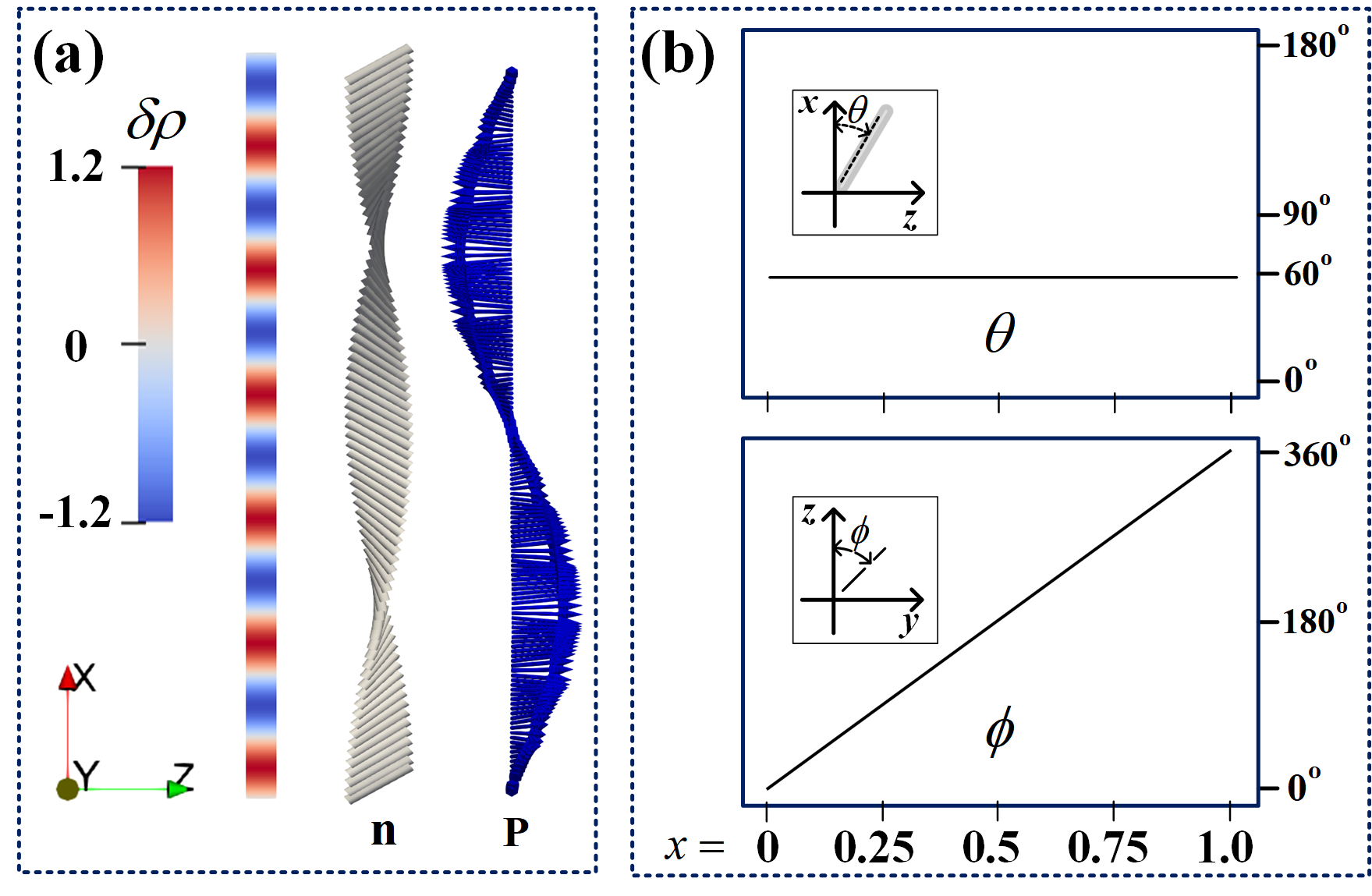}
    \caption{The profile of helix solution with parameters $q_0=12\pi$, $\theta_0=\pi/3$, $a=-5$, $b=5$, $c=10$, $d=-10$, $e=0$, $f=10$, $\eta=2.5$, $\lambda_1=\lambda_2=10^{-5}$.
    (a) The layer structure is represented by the density variation $\delta\rho$ with colormap from blue to red. The director $\mathbf{n}$ (gray rods) is the eigenvector corresponding to the largest eigenvalue of $\mathbf{Q}$. The polarisation $\mathbf{P}$ (blue arrow) is derived from $\mathbf{P} = -\nabla\cdot\mathbf{Q}$. In the following figures, the same representation is adopted for $\delta\rho$, the director $\mathbf{n}$, and the polarisation $\mathbf{P}$, and the same colorbar for $\delta\rho$.
    (b) The plots of tilting angle $\theta$ and rotating angle $\phi$ vs $x$.
    The tilting angle of directors $\mathbf{n}$ from the $x$-axis is $\theta = \arccos(n_1)$.  The rotation angle on the $yz$-plane is calculated by $\phi=\arctan\frac{n_2}{n_3}$ if $\arctan\frac{n_2}{n_3}\geq 0$, otherwise, $\phi=2\pi + \arctan\frac{n_2}{n_3}$, where $\mathbf{n}=[n_1,n_2,n_3]^\top$.
   }
    \label{fig:cstar}
\end{figure}

A helix solution $(\Qvec,\delta\rho)$, shown in Fig.~\ref{fig:cstar}, is found as a minimiser of the proposed non-dimensionalised free energy \eqref{eq:energy-3d}.
With the chosen layer wavenumber
$q_0=12\pi$, the system therefore contains six layers. Meanwhile, as $\sigma_0 = 2\pi$, the director $\mathbf{n}$ derived from $\mathbf{Q}$ undergoes a full pitch, during which the rotating angle $\phi$ linearly increases from $0$ to 
$2\pi$, and the tilt angle $\theta$ remains almost as a constant $\theta \approx \theta_0$. The flexoelectric polarisation, $\mathbf{P}$ is important due to the strong distortion of the director in SmC$^*$. This feature differs from SmA and SmC phases, as there is no distortions in their ground states. Here, $\mathbf{P}$ is calculated by $-
\nabla\cdot\mathbf{Q}$, which comes from \cite{pajkak2018nematic}. 
At $x=0$, the polarisation vector $\mathbf{P}$ is aligned with $\mathbf{e}_y=(0,1,0)^\top$. As $x$ increases from $0$ to $1$, 
$\mathbf{P}$ undergoes a clockwise rotation through an angle of $2\pi$.
The overall polarisation cancels out, and the system exhibits zero net polarisation.

\textit{(B) Bistable switching}---The SmC$^*$ phase composed of chiral molecules exhibits ferroelectricity.
FLCs are distinguished from other liquid crystal phases by the presence of a permanent dipole moment, which enables electro-optical response speeds 2-3 orders of magnitude faster than those of conventional nematic LCs.

We demonstrate such a crucial property by using the ferroelectric formulation, which is a coupling energy density $f_{elec}$ between the $\Qvec$-tensor and electric field $\mathbf{E}$:
\begin{equation}\label{eq:electric}
f_{elec} = e_1 \mathbf{E}\cdot(\nabla\cdot \mathbf{Q})-e_2 \mathbf{E}^T\mathbf{Q}\mathbf{E},
\end{equation}
where $e_1$ corresponds to the Meyer's flexoeletric coefficients, $e_2>0$ combines the permittivity constant of the vacuum for the applied electric field and the dielectric anisotropy of the material \cite{meyer2018electric}, and $\mathbf{E}$ represents the electric field.
The first term in \eqref{eq:electric} corresponds to the flexoelectric polarisation $-e_1\mathbf{E}\cdot\mathbf{P}$, and here we replace the representation of the polarisation vector $\mathbf{P}$ by $-
\nabla\cdot\mathbf{Q}$, the idea of which comes from \cite{pajkak2018nematic}. The second term is the dielectric term, which always dominates.

\begin{figure}
    \includegraphics[width=0.98\columnwidth]{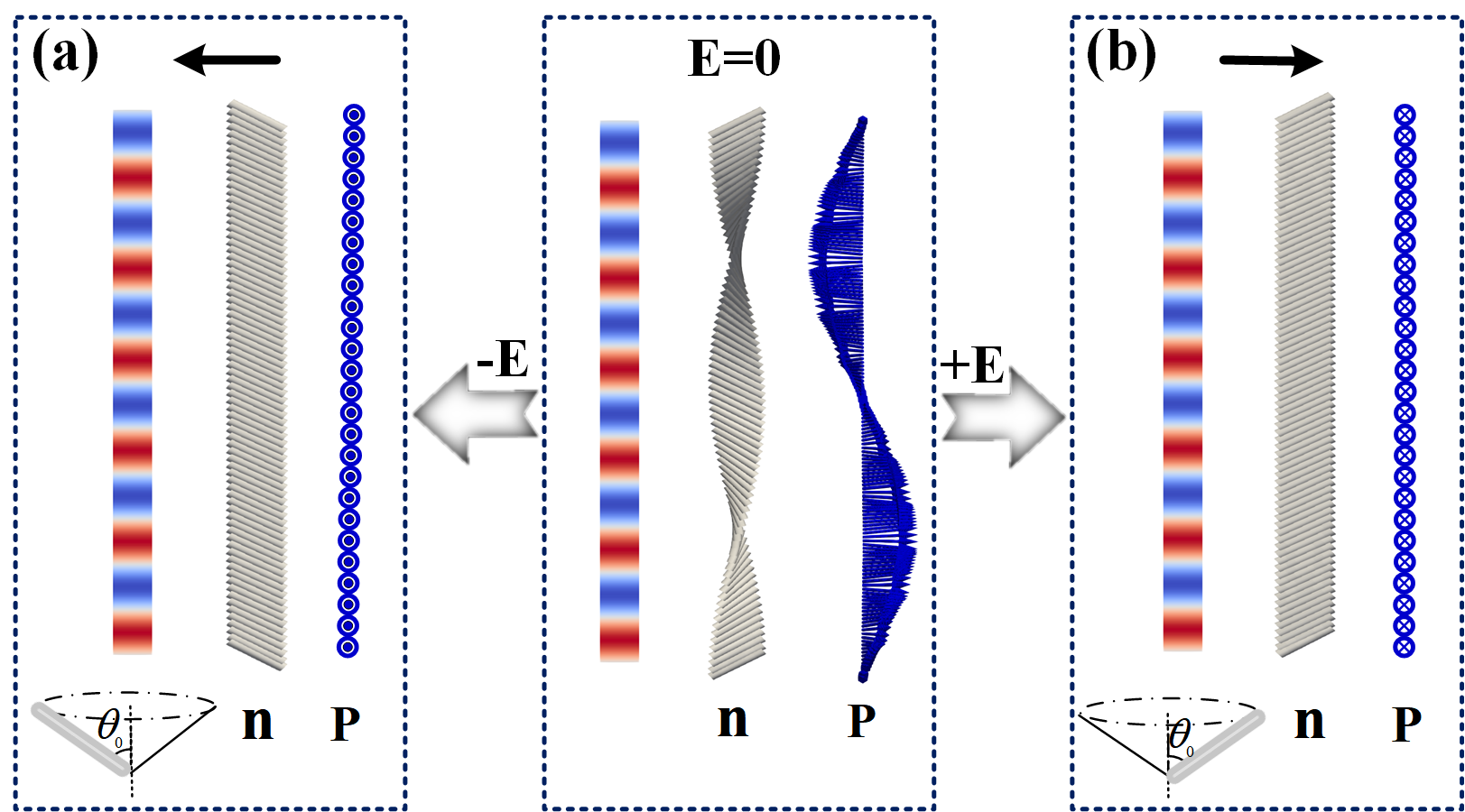}
    \caption{The bistable switching FLCs system with parameters $q_0=12\pi$, $a=-10$, $b=10$, $c=20$, $d=-10$, $e=0$, $f=10$, $\eta=1$, $\lambda_1=\lambda_2=10^{-4}$, $\theta_0=\pi/3$, $e_1 = 0.1$, $e_2=0.8$, and $\sigma_0=2\pi$. (a-b) The unwound states, the minimisers of the sum of the energy in \eqref{eq:energy-3d} and the integral of electric energy density in \eqref{eq:electric} with external electric field (a) $\mathbf{E} = (0,0,-1)^{\top}$ and (b) $\mathbf{E} = (0,0,1)^{\top}$.
   }
    \label{fig:elec-cstar}
\end{figure}

As reported in \cite{chen-2024NSR,yoshizawa2024,song2024giant}, the application of an external electric field leads to the emergence of a bistable ferroelectric effect in the SmC$^*$ phase (Fig. \ref{fig:elec-cstar}).
As we discussed in (A), without the application of an electric field, we obtain a helix solution without net polarisation. Applying the electric field perpendicular to layer normal, e.g., $\mathbf{E}=[0,0,w]^\top$ with $w>0$, molecules synchronously rotate towards the right side of the orientation cone (Fig. \ref{fig:elec-cstar}(b)). As the helix is unwound, there is no longer a bend distortion and $\mathbf{P}$ vanishes. By alternating the direction of the applied electric field with $w<0$, molecules respond to such a switch from one side to the opposite side of the cone surface (Fig. \ref{fig:elec-cstar}(a)). Periodic boundary conditions are applied in this scenario. 


\textit{(C) Bookshelf and Chevron}---The bookshelf structure, where the layers are stacked uniformly, and chevron structure \cite{gennes-book}, where layers adopt a characteristic V-shaped, are common structures observed in smectic phases, as also discussed in \cite{xia2024prr}. Therefore, a natural question is whether structures analogous to the bookshelf or chevron configurations may also arise in the SmC$^*$ phase.
Since the chevron structure is inhomogeneous in at least two dimensions, we carry out our study on a two-dimensional domain. We take the domain $\Omega = [0,1]^2$ in this scenario. 

\begin{figure}
    \includegraphics[width=1\columnwidth]{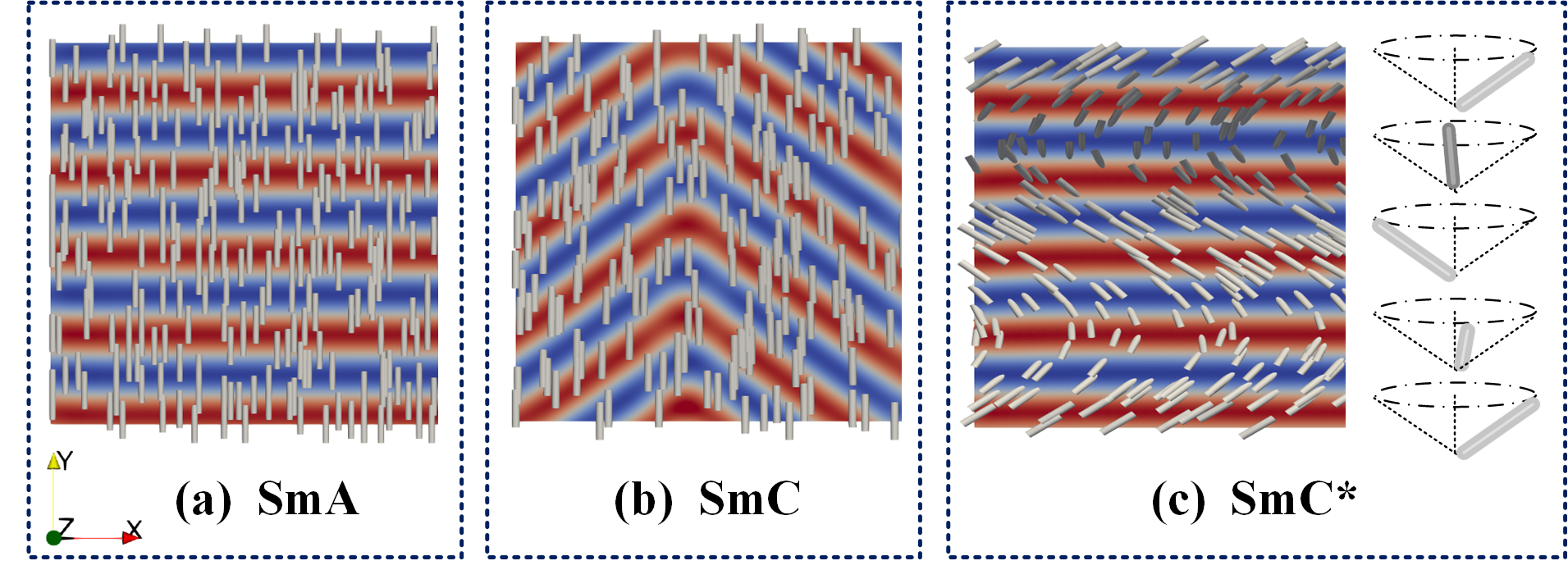}
    \caption{The profiles of minimisers of \eqref{eq:energy-3d} with parameters $q_0=10\pi$, $a=-30$, $b=30$, $c=60$, $d=-10$, $e=0$, $f=10$, $\eta=1$, $\lambda_1=\lambda_2=10^{-5}$. (a) The bookshelf structure in SmA phase ($\theta_0 = 0$, $\sigma_0 = 0$). (b) The chevron structure in SmC phase ($\theta_0 = \pi/3$, $\sigma_0 = 0$). (c) The helix structure in SmC$^*$ phase ($\theta_0 = \pi/3$, $\sigma_0 = 10\pi$). 
   }
    \label{fig:rectangle-cstar}
\end{figure}

Our proposed model \eqref{eq:energy_LdG} is capable of describing multiple smectic phases under various parameter settings.
When $\theta_0=0$ and $\sigma=0$, the model describes the simple SmA phase where molecules are parallel to layer normals. In Fig.~\ref{fig:rectangle-cstar}(a), directors are vertical-aligned in the present $xy$-domain due to the imposed vertical boundary condition of directors, thus leading to the bookshelf configuration.
When $\theta_0\neq 0$ and $\sigma = 0$, smectic layers tend to bend into the chevron configuration to accommodate themselves under the SmC setting, where molecules tilt over layer normals (Fig.~\ref{fig:rectangle-cstar}(b)).
When $\theta_0\neq 0$ and $\sigma \neq 0$, it comes to the SmC$^*$ phase where a cone-surface rotation should be respected, bookself-like uniformly stacked layers are observed.
Indeed, the profile in Fig.~\ref{fig:rectangle-cstar}(c) can be considered as a two-dimensional extension of the one-dimensional helix in Fig.~\ref{fig:cstar}.

The chevron also exists in SmC$^*$ and was recognised in 1987 by Clark and his collaborators by X-ray scattering experiments \cite{rieker1987chevron}.
The origin of chevrons in both SmC and SmC$^*$ phases is the same: the contraction of the smectic layer spacing upon molecular tilt, which can be relaxed through a folding instability of the layers. As far as we know, due to the well-known Surface-Stabilised Ferroelectric Liquid Crystal (SSFLC) \cite{ClarkLagerwall1980} systems, almost all of the related literature \cite{largewall-giesselmann2005,jones2012bistable,clark1988smectic} has focused on samples with sufficiently thin cell gaps or a rubbing direction to unwind the helical structure of the SmC$^*$ phase, reducing it to the case of the achiral SmC phase.

\textit{(D) Defect}---Another ubiquitous phenomenon concerned in the study of LCs is the appearance of defect structures. In \cite{chen-2024NSR}, the authors proposed a fast-switchable scheme based on the ferroelectric liquid crystal
topological structures, like $+1/2$- and $+3/2$-defect, to achieve the edge detection, which is a fundamental operation for feature extraction in image processing. The formation of focal conic domains (FCDs) of SmC$^*$ is investigated experimentally in \cite{choudhary2022hierarchical}.
A three–dimensional FCD configuration can be obtained by reconstructing from the profile on a disc with a 
$+1$ point defect at its center.\cite{stewart2019static}.

\begin{figure}
    \includegraphics[width=1\columnwidth]{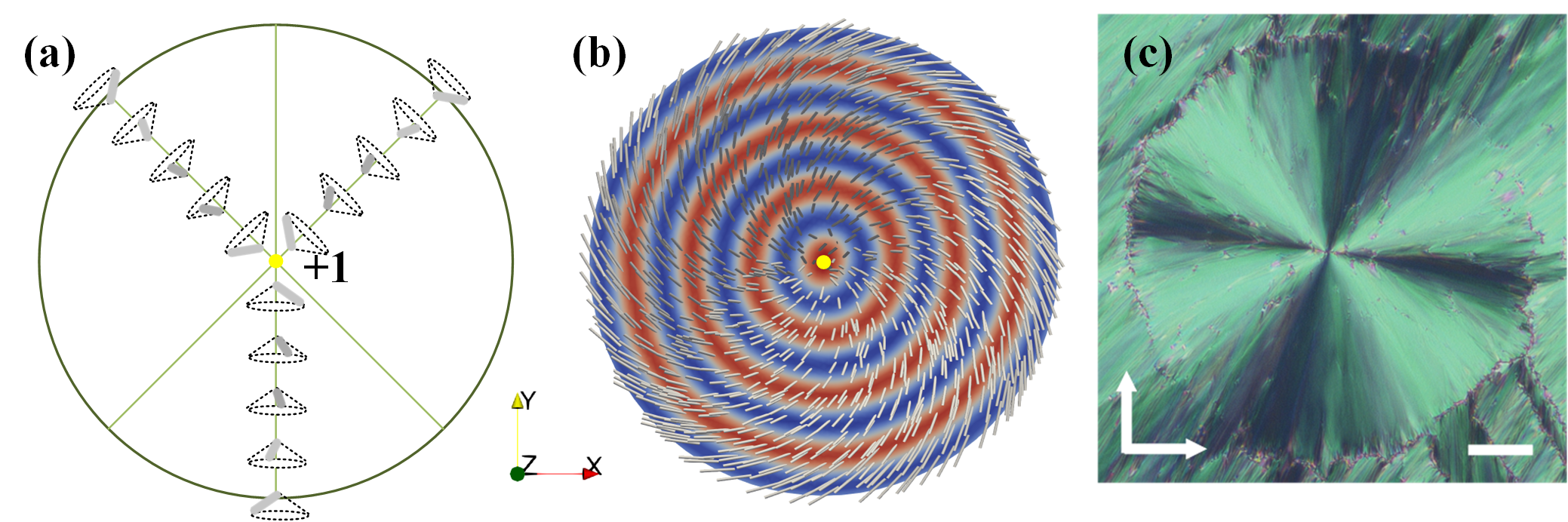}
    \caption{(a) Schematic illustration of directors around a $+1$ defect (yellow dot). (b) The profiles of the minimiser $(\mathbf{Q},\delta \rho)$ with a central $+1$ point defect on a unit disc, using parameters $q_0=8\pi$, $a=-10$, $b=10$, $c=20$, $d=-10$, $e=0$, $f=10$, $\eta=1$, $\lambda_1=\lambda_2=10^{-5}$, $\theta_0=\pi/3$, and $\sigma_0=\pi$.  (c) The POM image of a $+1$ defect texture in SmC$^*$. Scale bar: 5$\mu m$.
   }
    \label{fig:circ-cstar}
\end{figure}

For the sake of mathematical simplicity, we numerically reproduced the solution corresponding to a $+1$ defect in the SmC$^*$ phase.
To construct the initial guess of the $\Qvec$-tensor with a $+1$-defect at the origin, we assume $\mathbf{Q}_0$ is uniaxial, i.e., 
\begin{equation*}\mathbf{Q}_0 = s^*(\mathbf{n}_0\otimes\mathbf{n}_0-\frac{\mathbf{I}}{3})
\end{equation*}
where $s^*$ is defined in \eqref{eq:s*} and $\mathbf{n}_0$ is given by
\begin{equation*}
\mathbf{n}_0 = \mathbf{R}_z\mathbf{n}_{y0} = \begin{bmatrix}
    \cos(\theta)\cos(\theta_0)-\sin(\theta)\sin(\theta_0)\cos(\sigma_0r)\\
     \sin(\theta)\cos(\theta_0) + \cos(\theta)\sin(\theta_0)\cos(\sigma_0r)\\
     \sin(\theta_0)\sin(\sigma_0 r)
\end{bmatrix},
\end{equation*}
where $\theta = arctan(y/x)$ and $r = \sqrt{x^2+y^2}$.
The director of the helical structure on $y = 0$ is
\[\mathbf{n}_{y0}=(\cos(\theta_0),\sin(\theta_0)\cos(\sigma_0r),\sin(\theta_0)\sin(\sigma_0r)),\]
and the rotation matrix is
\[\mathbf{R}_z = \begin{bmatrix}
    \cos(\theta) &-\sin(\theta) &0\\
    \sin(\theta) &\cos(\theta) &0\\
    0 &0 &1
\end{bmatrix},\] which rotates directors counter-clockwise an angle of $\theta$ about the $z$-axis.
The Dirichlet boundary condition $\Qvec=\Qvec_0\vert_{r = 1}$ on $r = 1$ is imposed. The initial condition of $\delta\rho$ is 
\begin{equation}
\delta\rho_0 = \cos (q(\cos(\theta)x + \sin(\theta)y)).
\end{equation}
By minimising the free energy, we obtain the solution profile illustrated in Fig.~\ref{fig:circ-cstar}(b). The molecules rotate along the radial direction with the specified fixed angle $\theta_0$ of the cone surface.

Although the conditions do not fully correspond to those employed in the numerical simulations, we reproduced the corresponding configurations experimentally.
In the experiment, two ITO glass substrates covered with photoalignment agent SD1 are assembled together to give planar anchoring, and the cell gap is $4 \mu m$ \cite{wu2023topological}. A $+1$ alignment singularity pattern is recorded into the SD1 layer using a multistep partly overlapping photoalignment, and the alignment distribution can be expressed as: $\alpha(x,y) = \arctan \frac{y}{x}$. Then, the FLC materials (FD4004N, DIC Co., Japan, the helix pitch $p_0 = 1.0 \mu m$, spontaneous polarization $P_s = 10 nC/cm^2$, and tilt angle $\theta_0 = 22.5^\circ$, the phase transitions of isotropic state to chiral smectic A (SmA$^*$) and SmA$^*$ to SmC$^*$ are at $72^\circ$ and $65^\circ$, respectively) is filled into the prepared cell at $85^\circ C$ and then gradually cooled from isotropic phase to SmC$^*$ phase at a rate of $1^\circ C/min$. As a result, a Maltese cross is observed under polarised optical micrograph (POM) (see Fig.~\ref{fig:circ-cstar}(c)), indicating the generation of the $+1$ defect LC texture. In this case, the radial space-variant intrinsic defect lines imply the emergence of layered structures, and the spontaneous helices induce a slight in-plane rotation of the $+1$ defect.

\section{Conclusions}
In this paper, we propose a continuum tensorial model that can describe the chiral smectic C (SmC$^*$) liquid crystal phase, which is an extension of the model for SmA and SmC in \cite{xia2024prr,xia-2021-article} by adding the twisting elastic energy density. The model has two variables: a $\Qvec$ tensor measuring the orientational order of the molecules, and a real variable $\delta\rho$ for the smectic layer. The effect of external electric field on SmC$^*$ structures is modeled by adding the coupling energy between the order parameter $\Qvec$ and electric field $\mathbf{E}$, under the assumption that the natural ferroelectricity property of SmC$^*$ phases, the spontaneous polarisation $\mathbf{P}$ can be represented by $-\nabla\cdot\mathbf{Q}$. We
then illustrate our extension  through a series of numerical experiments, beginning with the one-dimensional helical structure and corresponding spontaneous polarisation, and bistable switching induced by external electric field. 
We then extend the study to a two-dimensional domain, comparing the equilibrium configurations of SmA, SmC, and SmC$^*$ phases within the same theoretical framework. Finally, we simulate the structure of a $+1$-charge defect in the SmC$^*$ phase on a unit disc, revealing the interplay between topological defects and helical twisting.  

The present model establishes a foundation for studying more complex SmC$^{*}$ structures, such as three-dimensional FCDs with dislocations \cite{choudhary2022hierarchical}, as well as the behaviors of different ferroelectric liquid crystal systems, including SSFLC, deformed helix ferroelectric liquid crystal (DHFLC), and electrically suppressed helix ferroelectric liquid crystal (ESHFLC), etc.~\cite{blinov-book}. 

Moreover, it provides a unified framework for investigating phase transition processes between distinct liquid crystal phases: isotropic, nematic, SmA, SmC and SmC$^*$ phases, whether driven thermotropically or lyotropically, under homogeneous conditions or in the presence of defects induced by confinement. Nevertheless, the explicit dependence of the model parameters on temperature and concentration remains an important subject for future study.

\textit{Acknowledgments}--The work of J.X.~is supported by the Hunan Provincial Natural Science Fund for Excellent Youths (Grant No.~2023JJ20046), and Young Elite Scientists Sponsorship Program by CAST (Grant No.~2023-JCJQ-QT-049), National Natural Science Foundation of China (Grant No.~12201636), and Science and Technology Innovation Program of Hunan Province (Grant No.~2023RC3013). J.W. is supported by the National Key Research and Development Program of China (2022YFA1203700), the National Natural Science Foundation of China (NSFC) (62035008 and 13001281).

\bibliography{ref}
\end{document}